\documentclass[useAMS,usenatbib]{mn2e}
\usepackage{epsfig,amsmath,amssymb,amsfonts,mathrsfs,latexsym,graphicx}
\include{journals}
\newcommand{\eg}{e.g.\ }
\newcommand{\ie}{i.e.\ }
\newcommand{\Msun}{M_{\odot}}

\newcommand{\kms}{km~s$^{-1}$}

\newcommand{\HeI}{He~{\sc i}}
\newcommand{\CI}{C~{\sc i}}

\newcommand{\OI}{O~{\sc i}}
\newcommand{\OII}{O~{\sc ii}}
\newcommand{\NaI}{Na~{\sc i}}
\newcommand{\MgII}{Mg~{\sc ii}}
\newcommand{\MgI}{Mg~{\sc i}}
\newcommand{\SI}{S~{\sc i}}
\newcommand{\SII}{S~{\sc ii}}
\newcommand{\SiI}{Si~{\sc i}}
\newcommand{\SiII}{Si~{\sc ii}}

\newcommand{\CaI}{Ca~{\sc i}}
\newcommand{\CaII}{Ca~{\sc ii}}

\newcommand{\FeI}{Fe~{\sc i}}
\newcommand{\FeII}{Fe~{\sc ii}}
\newcommand{\FeIII}{Fe~{\sc iii}}

\newcommand{\CoII}{Co~{\sc ii}}
\newcommand{\CoIII}{Co~{\sc iii}}

\newcommand{\NiIII}{Ni~{\sc iii}}
\newcommand{\Fefs}{$^{56}$Fe}
\newcommand{\Cofs}{$^{56}$Co}
\newcommand{\Nifs}{$^{56}$Ni}

\def\gsim{\mathrel{\rlap{\lower 4pt \hbox{\hskip 1pt $\sim$}}\raise 1pt \hbox {$>$}}}
\def\lsim{\mathrel{\rlap{\lower 4pt \hbox{\hskip 1pt $\sim$}}\raise 1pt \hbox {$<$}}}

\title[The ejecta of the Type Ic SN 2007gr]{
The Type Ic SN 2007gr: a census of the ejecta from late-time optical-infrared
spectra}

\author[P.A. Mazzali et al.]{Paolo A. Mazzali$^{1,2,3}$\thanks{E-mail: 
mazzali@mpa-garching.mpg.de}, I. Maurer$^{1}$, S. Valenti$^{4}$, R. Kotak$^{4}$, 
D. Hunter$^{4}$ \\
\\
$^{1}$Max-Planck Institut f\"ur Astrophysik, Karl-Schwarzschildstr. 1, D-85748 
Garching, Germany \\
$^{2}$Scuola Normale Superiore, Piazza dei Cavalieri, 7, 56126 Pisa, Italy \\
$^{3}$INAF-Osservatorio Astronomico, vicolo dell'Osservatorio, 5, I-35122 
Padova, Italy \\
$^{4}$Astrophysics Research Centre, School of Mathematics and Physics, 
Queen's University, Belfast, Belfast BT7 1NN, UK
}
\begin{document}

\date{Accepted ... Received ...; in original form ...}

\pagerange{\pageref{firstpage}--\pageref{lastpage}} \pubyear{2010}

\maketitle

\label{firstpage}

\begin{abstract}
Nebular spectra of Supernovae (SNe) offer an unimpeded view of the inner region of
the ejecta, where most nucleosynthesis takes place. Optical spectra cover most,
but not all of the emitting elements, and therefore offer only a partial view of
the products of the explosion. Simultaneous optical-infrared spectra, on the
other hand, contain emission lines of all important elements, from C and O
through to the Intermediate Mass Elements (IME) Mg, Si, S, Ca, and to Fe and
Ni.  In particular, Si and S are best seen in the IR. The availability of IR
data makes it possible to explore in greater detail the results of the
explosion. SN\,2007gr is the first Type Ic SN for which such data are available.
Modelling the spectra with a NLTE code reveals that the inner ejecta contain
$\sim 1 \Msun$ of material within a velocity of $\approx 4500$\,\kms. 
The same mass of \Nifs\ derived from the light curve peak ($0.076 \Msun$) was 
used to power the spectrum, yielding consistent results.  Oxygen is the dominant
element, contributing $\sim 0.8 \Msun$. The C/O ratio is $< 0.2$.  IME account
for $\sim 0.1 \Msun$. This confirms that SN\,2007gr was the explosion of a
low-mass CO core, probably the result of a star of main-sequence mass $\approx
15 \Msun$. The ratios of the \CaII\ lines, and of those of \FeII, are sensitive
to the assumed degree of clumping. In particular, the optical lines of [\FeII]
become stronger, relative  to the IR lines, for higher degrees of clumping. 
\end{abstract}


\section{Introduction}

Supernovae (SNe) are the final explosion marking the end of the life cycle of
different types of stars. Massive stars ($M \gsim 8 \Msun$) explode when their
core can no longer synthesise new elements and undergoes collapse. Thermonuclear
SNe (SNe Ia) are the complete explosive destruction of White Dwarfs reaching the
Chandrasekhar mass limit, most likely via accretion in a binary system
\citep{HillNie00}, although other channels have been proposed
\citep{Pakmor10,Fink10}. There are a number of ways to measure the properties of
SNe and infer those of the progenitor stars: classically, the analysis of SN
light curves yields information about the mass ejected in the explosion and the
explosion kinetic energy \citep{Arnett82}. Combined studies of the light curve
and the spectra allow a more precise estimate of these parameters, as well as of
the abundances in the ejecta, for both SNe\,Ia and core-collapse SNe 
\citep[\eg][respectively]{Mazz04eo,Mazz08D}. 

The properties of the progenitor star may be inferred through direct imaging
\citep[\eg][]{Smartt09} or searches for the surviving companion in the case of
SNe\,Ia \citep[\eg][]{PilarTycho}. However, the only method that looks directly
at the SN ejecta when they are transparent but not yet a remnant is
nebular-epoch spectroscopy. Soon after the explosion, the SN nebula is still
optically thick, and only the outer layers can be observed. About 6 months to
one year after the explosion, however, expansion makes the nebula sufficiently
thin that it becomes essentially transparent to radiation. In this phase the gas
is heated by collisions with the fast particles produced in the thermalization
process of the $\gamma$-rays and the positrons produced in the decay of \Cofs\
to \Fefs, and it cools via line emission. As optical depth effects are
negligible, the energy that is deposited is immediately re-radiated, which makes
it easier to estimate the \Nifs\ mass\ \citep{Axelrod80}. This phase can be used
to study the inner layers of SNe, yielding information on the details of the
explosion, both for SNe\,Ia \citep{MazzZorro} and core-collapse SNe, especially
those that have lost the outer layers \citep[Type Ib/c, ][]{Maeda08,Maurer10a}.

One major advantage of the nebular phase is that the luminosity emitted by a 
certain ion depends, albeit indirectly, on the mass of the ion itself. A reliable
estimate of the emitting mass requires then two basic elements: observational
coverage of the largest possible number of lines of different elements and a
reliable calculation of the ionization and excitation state of the gas. For the
latter, NLTE calculations are usually performed. The former requirement implies
essentially observing as wide a spectral range as possible. Traditionally,
because of the faintness of SNe in the nebular phase and the easier availability
of optical detectors, in almost all cases only optical spectra of SNe in the
nebular phase were obtained. However, several important line emission features
are located at IR wavelengths. This includes elements such as Si, which is very
important for SNe and has no strong emission lines in the optical, so that its
mass cannot be derived from optical spectra. Also, the ratios of different lines
of the same element can be used to determine the density of the emitting gas 
\citep[\eg \CaII,][]{LiMcCray93}, \citep[\FeII,][]{Leloudas09}. 

Only a very small number of nebular-phase IR spectra of SNe is available, and
these are typically SNe\,Ia spectra \citep[\eg ][]{Spyro04,Motohara06}. Among
core-collapse SN, there are late-time IR spectra for the peculiar SN\,IIP 1987A
\citep{Fassia02}, and for the highly reddened SN\,IIP 2002hh \citep{Pozzo06}.
The first stripped-envelope SN for which late-time IR spectra have been obtained
is the Type Ic SN\,2007gr \citep{Hunter09}. 

SNe Ic show neither H nor He in their spectra \citep{Fil97}, and come with a
very diverse range of observational \citep{Matheson00} and physical properties
\citep{Mazz03lw}. These depend on the properties of the progenitor star and the
details of the explosion, ranging from ejected masses of $\sim 1 \Msun$, \Nifs\
masses of $\sim 0.1 \Msun$ and kinetic energies of $\sim 10^{51}$\,erg for
``normal" events \citep[\eg][]{Sauer06} to ejected masses of $\sim 10 \Msun$,
\Nifs\ masses of $\sim 0.5 \Msun$ and kinetic energies of $\sim
5\cdot10^{51}$\,erg for GRB-connected hypernovae \citep{Mazz03lw}. 

SN\,2007gr had normal luminosity and spectral appearance, but was characterised
by the unusual presence of carbon lines in the early-time spectra, indicating
that C was not fully burned to O \citep{Valenti08}. A simple light curve
analysis \citep{Hunter09} suggests that SN\,2007gr ejected $\sim 2 \Msun$ of
material with kinetic energy $E_K \sim 1-4 \:10^{51}$\,erg, which would make it
a rather ordinary SN\,Ic. High-resolution radio observations of SN\,2007gr
revealed emission from an expanding source \citep[][but see
\citet{Sod10}]{Paragi10}.  This expansion was interpreted as evidence for
material moving at mildly relativistic velocities in what seems to be a bipolar
jet. This is at odds with the existing picture where all SNe associated with
emission from relativistic material also show large expansion velocities in the
early phase ($v \sim 0.1 c$), which links the relativistic material to the bulk
of the SN ejecta, and are characterised by large values of the ejected mass and
the explosion energy \citep{Mazz03lw}.

We use the nebular spectra of SN\,2007gr, including one simultaneous optical-IR
spectrum obtained more than one year after the explosion, to determine the
properties of the inner ejecta. 

In the following, in Section 2 we briefly review the data we use, in Section 3
we discuss the nebular model used for the calculations, Section 4 presents the
results of one-zone models, Section 5 deals with the models obtained using a
density profile, and in Section 6 our results are discussed.

\section{The Data} 

Nebular-phase spectra of SN\,2007gr were obtained at various epochs, as outlined
in \citet{Hunter09}. Earlier optical spectra have epochs of 103 to 158 days
after maximum. These spectra are not fully nebular. The same is true for various
IR spectra obtained at similar epochs. At day 375 after maximum a simultaneous
optical-IR spectrum was obtained. Since the optical and the IR spectra were
obtained within one day of one another, spectral evolution and relative
calibration are not an issue, unlike data at earlier epochs. The very advanced
phase of this later spectrum ensures that it is fully nebular. The spectrum was
presented in Figures 7 (optical) and 12 (IR) of \citet{Hunter09}. 

Here we use the very late optical-IR spectrum at 376.5 days after maximum for
our calculations\footnote{This spectrum was incorrectly labelled as +375.5 days
in \cite{Hunter09}, who did not take into account that 2008 was a leap year.},
and verify results using the optical spectrum at 158 days after maximum.

SN\,2007gr exploded in a crowded region in the spiral galaxy NGC 1058. In
particular, SN\,2007gr is flanked on either side by a relatively bright stellar
association \citep[see][Fig. 1]{Hunter09} and several fainter sources within 150
pc \citep[see][Fig. 1]{Crockett08}. At the late epochs under consideration, when
the SN is faint in comparison with these adjacent objects, particular attention
must be paid to the estimation of the background flux. In order to estimate the
contribution of these sources we extracted spectra obtained in the vicinity of
SN\,2007gr. The resulting background contribution clearly depends on the choice
of location at which the flux was estimated. We experimented with several
extractions along the spatial profile and chose the background contribution that
best matched the underlying slope in the SN continuum. We then scaled this
background to the apparent SN continuum and subtracted it from the SN spectrum
(no strong continuum is expected from the SN at late epochs). The contaminating
sources (within 150 pc) are probably clusters \citep{Crockett08} with spectral
energy distributions that peak at blue wavelengths, so while some contamination
affects the entire spectrum, it is expected to be particularly severe at bluer
wavelengths.

\section{The model} 

We modelled the spectra of SN\,2007gr using a code which computes the energy
deposition from the radioactive decay of \Nifs\ and \Cofs\ in the SN ejecta,
uses the deposited energy to estimate ionization and excitation, and balances
this with gas cooling via line emission in a NLTE scheme \citep{RLL92}. The
method follows that developed by \citet{Axelrod80}. 

The code allows the use of clumping, expressed in terms of a clumping factor
$\zeta$, which is defined as the inverse of the filling factor
\citep{LiMcCray93,Maurer10a}. This assumes that the gas is distributed
uniformly, but resides in clumps. The effect of clumps is to increase the local
density, and hence to favour recombination. Clumping also reduces the population
of the excited levels, which may be important in the context of the ratio of
optical to IR flux, as suggested by \citet{Leloudas09}. For SNe\,Ib/c,
\citet{Mazzali01} found that a typical clumping  $\zeta = 10$ yields best
results, while for SNe\,Ia clumping is not regarded to be necessary if a
detailed model of the explosion is used \citep[\eg][]{Mazz04eo}. The need for
clumping in SNe\,Ib/c may be the result of an inaccurate treatment of
ionization.  

Improvements in the treatment of ionization \citep{Maurer10b} make the code more
reliable, in particular for SNe where Ni and Fe do not dominate the ejecta. 
These improvements were introduced in the context of including a treatment of
oxygen recombination. A simultaneous improvement of the treatment of ionization
was also necessary in order to make the calculation of the recombination rate
more reliable. Forbidden line emission is influenced by this new treatment only 
weakly. Results with the old and new codes are shown and compared in Section 4.

In a one-zone version, the code assumes that the gas is distributed with uniform
density and abundances within a sphere of radius $R = v \times t$, where $v$ is
the outer velocity of the expanding nebula, as determined from the width of the
emission lines, and $t$ is the time elapsed from explosion. This takes advantage
of the fact that in SN ejecta any given parcel of matter is characterised by a
velocity which remains constant as long as the ejecta are not slowed down by
interaction with circumstellar material. Results obtained with this model are
presented in Section 4. 

In a more sophisticated version the SN envelope is divided into uniform radial
shells. Deposition of $\gamma$-rays and positrons is computed using a Montecarlo
scheme, and emission is computed within each shell. This version can be used if
a detailed model of the explosion is available, which describes the run of
density with ejecta velocity, or if one is sought for from fitting the profile
of the nebular lines \citep[\eg][]{Mazzali02apneb}. Results obtained with this
model are presented in Section 5. 

For the distance and reddening to SN\,2007gr we use the same values (d=9.3\,Mpc,
\ie distance modulus $\mu=29.84$\,mag; E(B-V)$=0.092$) as in \citet{Hunter09}.

\section{One-zone models} 

For the nebular spectra of SN\,2007gr we used first the 1-zone code to define
the general properties of the ejecta. Here we show 1-zone model results obtained
first without and then with the oxygen recombination module and compare them. 
We then applied the shell model to improve the results and the fits. These are
discussed in the next section.

\subsection{One-zone models, old code}

\subsubsection{Day 172} 

The first nebular spectrum we model was obtained on 2008 Jan 29. We assign it an
epoch of 172 days after explosion, using an arbitrary risetime of 13.5 days,
which is typical for SNe\,Ic and is consistent with the risetime deduced by
\citep{Hunter09}, $11.5 \pm 2.7$ days. 

The strongest lines in the spectrum are typical of SNe\,Ic: [\OI] 6300,
6363\,\AA, \MgI] 4570\,\AA, \CaII] 7291, 7324\,\AA, the \CaII\ IR triplet near
8600\,\AA, and \NaI\ D near 5900\,\AA. \CaII\ H\&K may be seen at the blue edge
of the spectrum. The spectrum can be fitted reasonably satisfactorily assuming
an outer velocity of 5000\,\kms\ and $\zeta = 10$, which is typical for SNe\,Ic
(Figure 1, blue line). The \Nifs\ mass was kept fixed at $0.076 \Msun$, as
indicated by the light curve peak \citep{Hunter09}. This value yields fairly 
good results. The mass enclosed within 5000\,\kms\ is $\sim 1.4 \Msun$, oxygen
being the main constituent ($\sim 1 \Msun$).

\begin{figure}
\includegraphics[width=89mm]{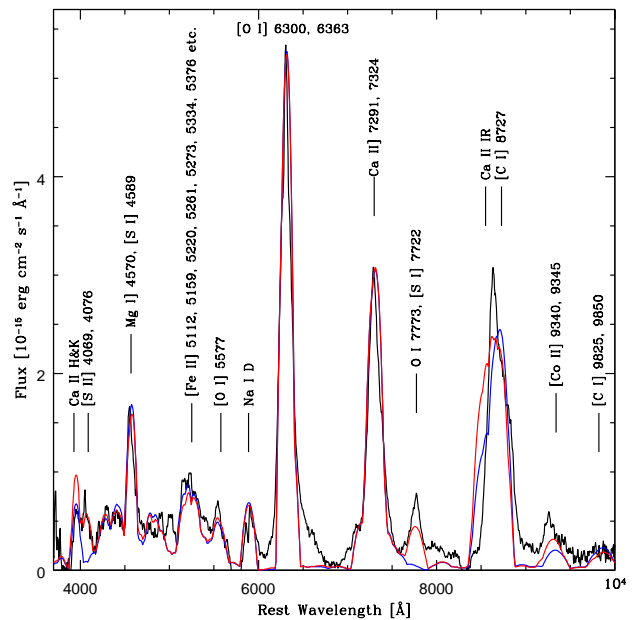}
\caption{SN\,2007gr: one-zone models for the spectrum at day 172 (29 Jan 2008,
black) with the old (blue) and the new version of the nebular code, which
includes an improved treatment of ionization and oxygen recombination 
(red). }
\label{d172_onezone}
\end{figure}

While most emission lines are reproduced, and identified in Figure 1, the main
shortcomings of the model are as follows. The line near 4050\AA\ is not 
reproduced. This is probably [\SII] 4069, 4076\,\AA, which is predicted to be
present but is much too weak when compared to the data. The low ionization
degree of sulphur ($\sim 5$\% in the models with the old treatment of
ionization) is probably responsible for the weakness of the emission line. The
broad base of [\OI] 6300, 6363\,\AA\ is not reproduced. This is the effect of
adopting a sharp outer boundary to the nebula. Shell models will improve this
(see Section 5). The emission line at $\sim 7700$\,\AA\ is not reproduced. This
is most likely the recombination line \OI\, 7773\,\AA, which is addressed in the
next sub-section. An alternative possibility is [\SI] 7723\,\AA. This is indeed
predicted by the model as the strongest line of sulphur, but is much too weak.
The second strongest sulphur line in the model is [\SII] 4068, 4076, which is 
too weak for a S mass of $0.08 \Msun$. The third strongest feature is [\SI]
4507, 4589\,\AA, which is completely swamped by the strong \MgI] 4570\,\AA\
line. In order to reproduce both the line near 7750\,\AA\ and that near 4050\AA\
as due to sulphur, the S mass must increase to an unreasonable $0.65 \Msun$,
confirming that the line near 7700\,\AA\ is \OI\ 7773. Additionally, cooling by
S emission weakens the other lines, so that all other masses must increase, to a
total of $\sim 2.4 \Msun$.  Such a large S mass fraction seems to contradict
nucleosynthesis results \citep{WW95}. 


The line at 8700\,\AA\ is too weak. This is a blend of \CaII\ IR triplet and
[\CI] 8727\,\AA. Increasing the strength of the [\CI] line is not an option, as
the wavelength of the line does not match that of the emission peak (assuming
spherical symmetry). Also, the strength of the other carbon line, [\CI] 9825,
9850\,\AA, is already overestimated. Therefore, a weak \CaII\ line is the likely
cause of the discrepancy. This may indicate a different degree of clumping than
adopted in the calculation \citep[see][]{LiMcCray93}. We however refrain from
investigating clumping here, since this may be better done using the optical/IR
spectrum of day 390, which has a broader wavelength coverage, and the new code,
which includes oxygen recombination and an updated treatment of ionization.
Finally, the line at 9300\,\AA, which is mostly due to [\CoII] 9340, 9345, is
not correctly reproduced.  This may be caused by insufficient background
subtraction.

\subsubsection{Day 390} 

The later spectrum is fully nebular, and it is therefore to be expected that 
our code can reproduce it better than the earlier one. Additionally, it covers
both the optical and the IR essentially simultaneously, making it possible to
narrow down the range of acceptable parameters. Many of the strongest features
are the same as on day 172. One of the key features for modelling is the [\FeII]
line near 5200\,\AA. This is composed mostly of lines at 5159, 5262, 5273, and
5333\,\AA. Other strong lines include the structured blend near 8700\,\AA, which
is due to a mix of a weak \CaII\ IR triplet, [\CI] 8727\,\AA, and [\FeII] lines,
the strongest being at 8617 and 8892\,\AA\ (Figure 2). The [\FeII] 8617\,\AA\
line is actually the strongest of all these emissions, which explains why the
observed profile peaks between \CaII\,IR and the [\CI] line.  Uncertainties in
the strength of the [\FeII] lines may be responsible for the poor reproduction
of the corresponding emission at day 172. 

\begin{figure}
\includegraphics[width=89mm]{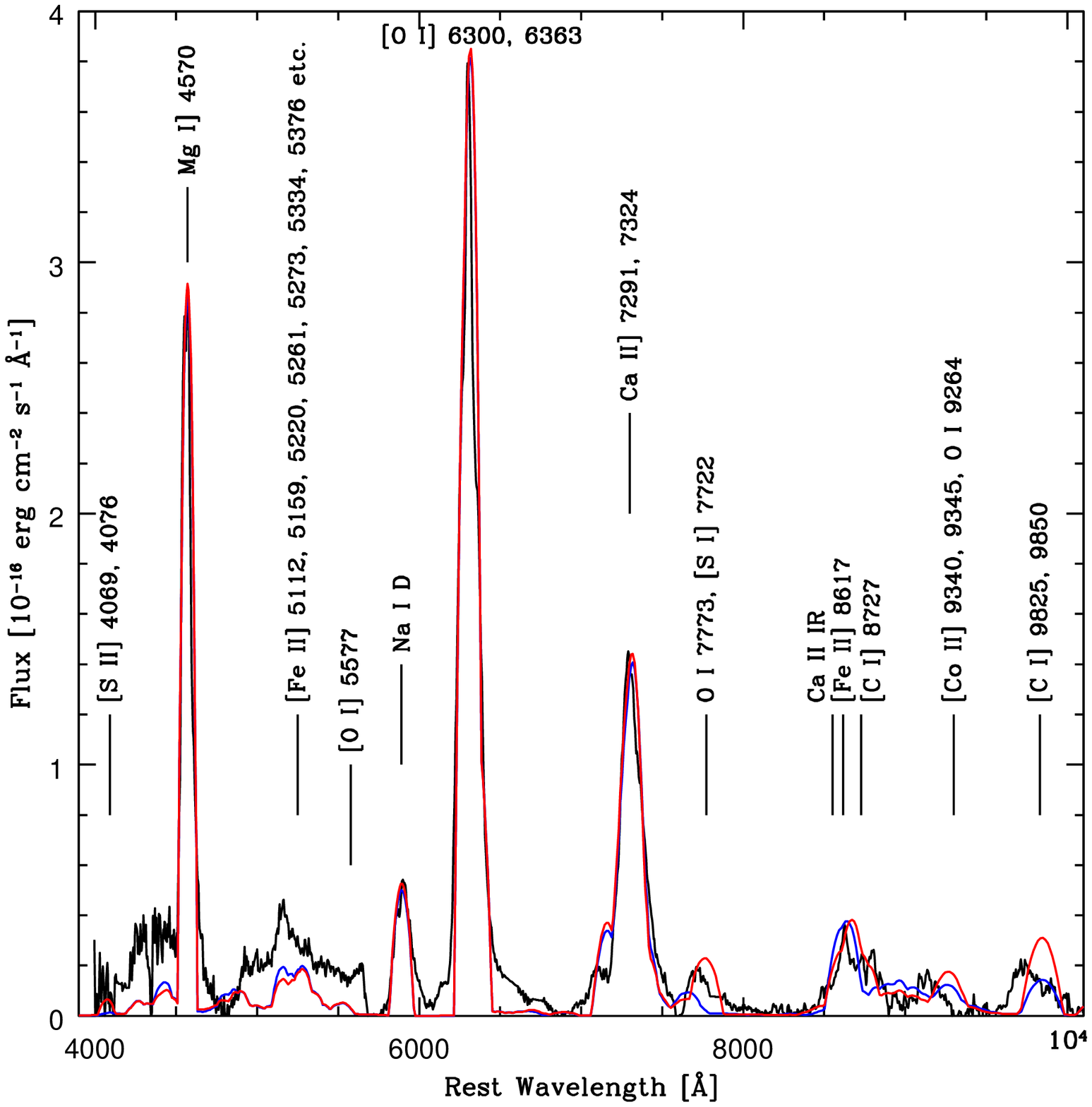}
\caption{SN\,2007gr: one-zone models for the spectrum at day 390 (3 Sept 2008,
black) with the old (blue) and the new version of the nebular code, which
includes an improved treatment of ionization and oxygen recombination 
(red). }
\label{d390_onezone_opt}
\end{figure}

\begin{figure}
\includegraphics[width=89mm]{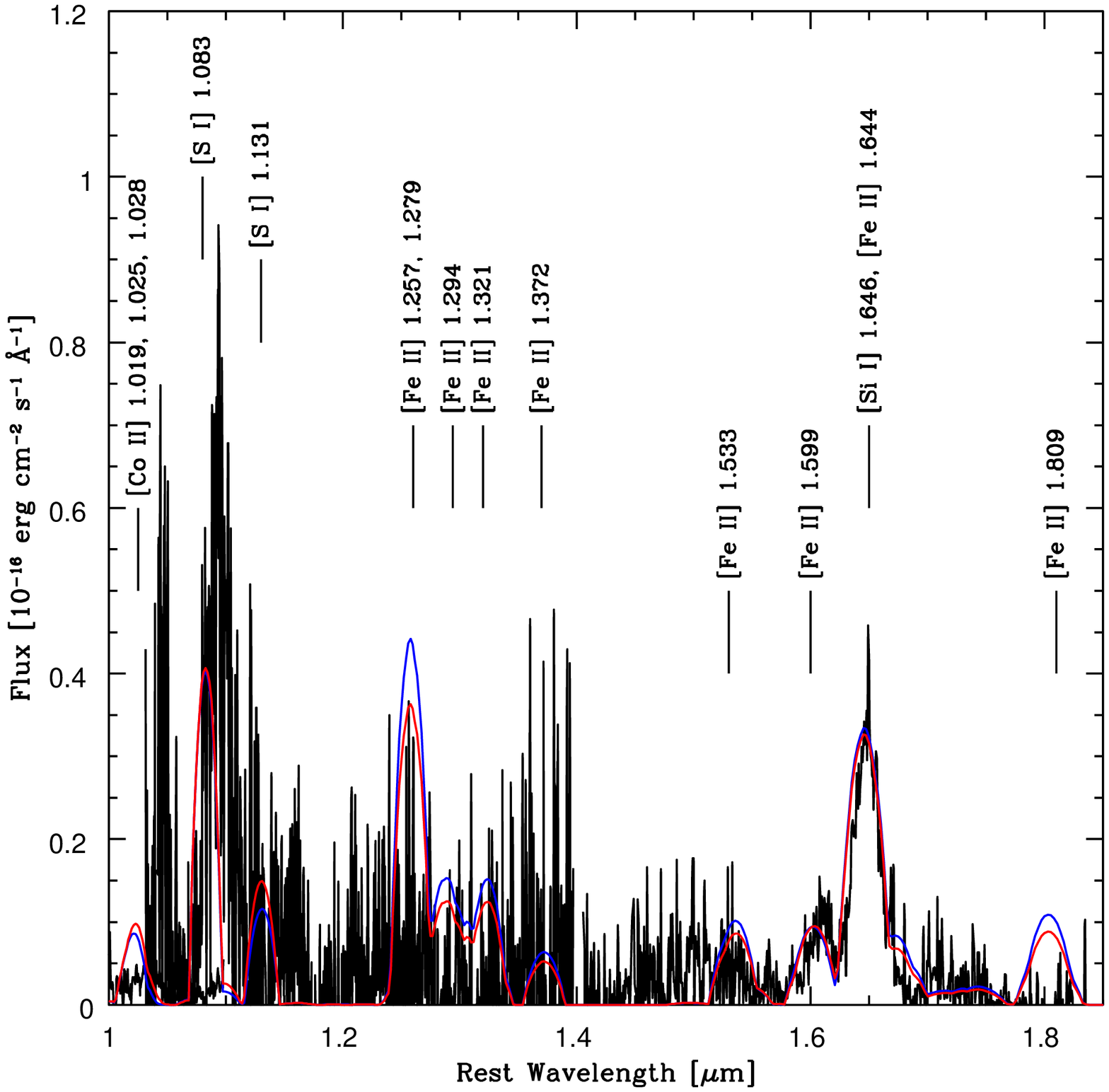}
\caption{SN\,2007gr: one-zone models for the spectrum at day 390 (3 Sept 2008,
black) with the old (blue) and the new version of the nebular code, which
includes an improved treatment of ionization and oxygen recombination 
(red). }
\label{d390_onezone_IR}
\end{figure}

Moving to the IR, several other features are visible which allow relative
comparisons. They are labelled in Figure 3. One of the most useful is [\SiI]
1.604, 1.646\,$\mu$, which is the strongest silicon line, blended with [\FeII]
1.599, 1.644, 1.664\,$\mu$. Lines of [\FeII] are also seen near 1.25\,$\mu$
(strongest contributors to the feature being lines at 1.249, 1.257, 1.270,
1.279, 1.321\,$\mu$), but the spectrum is very noisy there. Other lines of IME
are also present in the IR.

\begin{table*}
 \centering
 \begin{minipage}{170mm}
  \caption{Results of One-zone Models.}
  \begin{tabular}{cccccccccccccc}
  \hline
 code v. & epoch & vel & Mass & T$_e$ & log(n$_e$) & M(C) & M(O) & M(Ne) &  M(Mg) & M(Si) & M(S) & M(Ca) & M(\Nifs) \\
   	 &[days]& \kms& $\Msun$& K & cm$^{-3}$& $\Msun$& $\Msun$& $\Msun$& $\Msun$ & $\Msun$ & $\Msun$& $\Msun$ & $\Msun$ \\
 \hline
   old   &  172  & 5000& 1.40 & 5100 &  7.4 & 0.25 & 0.99 & 6.0e-5 & 1.25e-3 & 0.006 & 0.023 & 0.053 & 0.076 \\
   new   &  172  & 5000& 1.83 & 4870 &  7.7 & 0.25 & 1.40 & 6.0e-5 & 1.00e-3 & 0.006 & 0.070 & 0.019 & 0.076 \\
   old   &  390  & 4000& 1.04 & 3560 &  6.4 & 0.05 & 0.81 & 2.7e-4 & 2.80e-2 & 0.006 & 0.023 & 0.048 & 0.076 \\
   new   &  390  & 4000& 1.28 & 3430 &  6.6 & 0.14 & 0.99 & 3.5e-4 & 3.25e-2 & 0.009 & 0.024 & 0.013 & 0.076 \\
 \hline
\end{tabular}
\end{minipage}
\end{table*}

Since the flux in the IR lines of iron is basically determined from the flux in
the optical lines (despite some influence of clumping) the abundance of Si can
be determined from the line near 1.65\,$\mu$. This line has contributions from
both [\SiI] 1.608, 1.646\,$\mu$\ and [\FeII] 1.644, 1.664\,$\mu$ 
\citep{Hunter09}. Given the abundances adopted in our model, Fe and Si
contribute roughly equally to the observed emission. Fitting any difference
between the observations and the synthetic [\FeII] feature as due to [\SiII] is
actually the only way to derive the abundance of Si from the nebular spectra. 
The abundance of sulphur can be determined from the strong [\SI] line at
1.1\,$\mu$. This is composed of lines at 1.08 and 1.13\,$\mu$, which are much
stronger than the sulphur lines which are seen in the optical. Unfortunately the
region of the spectrum near 1.1\,$\mu$ is very noisy. The ratio of optical to IR
[\FeII] emission lines is affected by the degree of clumping in the ejecta.
However, since we have an independent estimate of the \Nifs\ mass from the light
curve, we can use that value ($0.076\Msun$), which basically fixes the amount of
iron at day 390, and modify the filling factor to obtain a best match to the
optical-IR Fe emission. Any additional flux in the 1.6\,$\mu$ line can then be
attributed to silicon, while at 1.1\,$\mu$ sulphur can be assumed to
contribute. 

Using the one-zone model, the spectrum can be best matched using an outer
velocity of $4000$\,\kms. This is significantly less than at day 172. Clearly
the outer regions of the nebula are becoming transparent, and the emission lines
become narrower \citep[see also][]{Tauben09}. This evolution indicates that the 
density structure in the ejecta is important in shaping the nebular emission
lines, and it suggests that the approximation of a one-zone model is not ideal.
As in the model for day 172, we used a \Nifs\ mass of $0.076 \Msun$. A clumping
factor $\zeta = 10$ yields the best ratio of optical to IR flux. At this epoch
the model contains a mass of $\approx 1 \Msun$, and it is again dominated by
oxygen ($\sim 0.8 \Msun$). Other elements have smaller mass. In particular, the
Si mass required to reproduce the $1.6 \mu$ line is only $0.006 \Msun$. This is
a very small mass, especially considering that other IME are more abundant (\eg\
S, Ca, Mg). Still, this model reproduces most features. The main shortcoming may
be in the [\FeII] line near 5200\,\AA, which may suggest that a somewhat larger
\Nifs\ mass may be required. However,  background subtraction may be uncertain
here, since the shape of the feature is reproduced reasonably well, as is the
ratio of other Fe lines in the optical and IR. Interestingly, this ratio
increases for increasing clumping factor. The higher density in the clumps
favours recombination and a greater role of collisional processes, which support
the population of the excited levels. Otherwise, at the low temperatures that
characterise the SN nebula at very advanced epochs, IR lines should dominate
\citep[the so-called IR catastrophe,][]{Axelrod80}. We present a model with more
clumping in Section 5. 

Although the mass of elements such as O and Ca is consistent at the two epochs,
others change significantly. In particular, the Mg mass in the later spectrum is
much larger than in the earlier one and the same is true, to a lesser extent,
for Na. Clearly these elements are not very well constrained by the model,
probably because of an incomplete treatment of ionization. Quite likely, both Mg
and Na are highly ionized at early epochs and become less ionized at later
phases. Given the low ionization potentials of \NaI\ and \MgI, the high
ionization in the early phase is probably caused by photoionization and possibly
thermal electron collisional ionization, neither of which are treated by the
code. An estimate of the masses of Na and Mg based on the neutral species at
times when ionization is still significant may therefore yield incorrectly large
values if this ionization is not taken into account.  Mass estimates obtained
around 390 days should be more reliable than at 172 days. 

Conversely, the C mass is much smaller in the later spectrum. Although this
estimate is uncertain also because of the poor fit of the 8600\,\AA\ emission,
to which [\CI] contributes, the reduction that is observed may indicate that C
is preferentially located at velocities between 4000 and 5000\,\kms. This is not
unreasonable given the structure of a stellar core: helium seems to "bottom out"
at $\sim 8000$\,\kms\ in SN\,2007gr, while C is observed at $v \sim
5000-8000$\,\kms\ also in the early phase \citep[][Fig. 14]{Hunter09}.

\subsection{One-zone models, updated code}

We have recently introduced a number of updates and improvements to the original
code. These are described by \citet{Maurer10b}. In particular, we have 
introduced recombination of \OI\ and revisited the treatment of ionization. The
former improvement was necessary in order to reproduce the \OI\ 7773\,\AA\
emission line, which was completely missing in the old version of the code
\citep[\eg][]{Mazzali02apneb}. The improved treatment of ionization was
necessary in order to provide a more accurate estimate of the electron density
so that recombination rates can be reliably computed.  While ionization rates
were obtained from a simple analytical form in the older version of the code, in
the new version they are calculated taking into account the atomic and
electronic loss processes of non-thermal electrons explicitly
\citep{Maurer10b}. 

Using the improved treatment of ionization \citep{Maurer10b}, oxygen and other
light elements are more ionized than they are using the old, simpler treatment,
while the ionization rates of Fe-group elements do not change significantly.
Therefore the ionization balance of SNe\,Ia (which are dominated by Fe-group
elements) is not influenced much by the new treatment of ionization, while the
ionization balance of SNe\,Ib/c, which are dominated by carbon and oxygen, is.

The increased ionization of oxygen and other light elements results in a larger
electron density, which in turn favours recombination and the formation of the
\OI\ 7773\,\AA\ line, as explained in \citet{Maurer10b}. In practice, this turns
out to reduce the need for clumping with respect to the old version of the code.
Using the old ionization treatment it was possible to mimic the effect of higher
ionization (higher electron densities) introducing a clumping factor. This can
explain why clumping seemed to be necessary in SNe\,Ib/c but not in SNe\,Ia.  As
an example, the spectra of SN\,2007gr can be reproduced using $\zeta = 2.5$.
This is sufficient to guarantee that \FeIII\ recombines to \FeII\ (in SNe\,Ib/c
[\FeIII] lines are not seen, unlike SNe\,Ia). In fact, with the new code the
\FeIII/\FeII\ ratio is similar to the value obtained with the old code. 
Although clumping is not necessarily required to obtain good fits to the
forbidden line observations, it may nevertheless be important e.g. for the
formation of the \OI\ 7773\,\AA\ line \citep{Maurer10b}. 

The abundances of most elements in SN\,2007gr (and other SNe\,Ib/c) are
determined from neutral or singly ionised species (\eg \CI, \OI, \MgI, \NaI,
\SiI, \SI, \FeII, \CoII). Higher ionization weakens the flux in these lines.
This must be offset by somewhat increasing the mass of these elements in the
nebula. Since the less ionized species remain the dominant ones even with the
updated treatment of ionization, the effect on the mass estimate is small. One
exception is calcium, for which the abundance is obtained from the more highly
ionised species, \CaII. In this case, higher ionization results in a smaller
calcium mass (by a factor of 2-4 depending on the individual case).

The results for day 172 and 390 are shown in Figures 1, 2, and 3 as red lines.
While the fits are similar to those obtained with the old model, the masses of
some elements change somewhat. We now discuss the two epochs in turn.

\subsubsection{Day 172} 

The new one-zone model for day 172 is characterised by  the same outer velocity
($5000$\,\kms) and \Nifs\ mass ($0.076 \Msun$) as the old code, and the results
are similar. However, as we discussed above, less clumping is required ($\zeta =
2.5$) since the ionization degree is higher. In the old model the average
ionization degree was $\sim 1/30$ (\OII/\OI, \FeIII/\FeII, etc.), the electron
density $n_e \approx 2.5\cdot 10^7$\,cm$^{-3}$ and the electron temperature $T_e
\approx 5100$\,K. With the improved treatment, the ionization of certain
elements (O, Mg, Si, S) is higher, $\sim 20$\%. The resulting increase in $n_e$
favours recombination of \FeIII, \CoIII\ and \NiIII, reducing the need for
clumping. The ionization of Ca increases less than for lighter elements, and it
is still only $\sim 5$\%, while that of Fe is essentially unchanged. The
electron density is now $n_e \approx 7 \times 10^7$\,cm$^{-3}$ and the electron
temperature $T_e \approx 4900$\,K. 

The increased fraction of highly ionized material weakens the flux in lines of
ions such as \OI, \MgI, \SiI, making it necessary to include more mass in the
ejecta. The total mass is now $\sim 1.8 \Msun$ within $5000$\,\kms. Oxygen is
still the most abundant element, with a mass of $\sim 1.4 \Msun$, followed by C
($0.25 \Msun$). The masses of the other elements are comparable to the old
model, since their degree of ionization does not change, except for Ca, which is
reduced by about a factor of 3 ($0.016 \Msun$), in accordance with the increase
in ionization. Calcium is the element for which the new treatment of ionization
has the biggest impact, as the Ca mass is estimated from the less abundant
singly ionised species. Iron-group lines are also mostly from singly ionised
species (\FeII, \CoII), but these are actually the highly abundant lower
ionization states. 

The improvements to the synthetic spectrum introduced by the updated model are
subtle, but one can see that the recombination line \OI\ 7773\,\AA\ is now
present (Figure 1, red line). The mismatch with respect to the data may be due
to background subtraction: the strength of the line seems to be reproduced
reasonably well. Alternatively, several \FeII\ lines near 7800\,\AA\ may be
poorly reproduced if their collision strengths are not accurate. With the higher
ionization, a weak [\SII] line is also visible near 4000\,\AA. 

The sharpness of the peaks of most emission lines contrasts with the more
rounded shapes of the synthetic profiles, which are assumed to be parabolae as
expected from a homogeneous, homologously expanding sphere. This can be improved
upon using a multi-zone approach.

\subsubsection{Day 390} 

As for the day 172 model, the outer velocity ($4000$\,\kms) and \Nifs\ mass
($0.076 \Msun$) used were the same with the new code as with the old one. The
same clumping as on day 172 ($\zeta = 2.5$) was used. In the optical, the main
improvement is that now \OI\ 7773\,\AA\ is reproduced (Figure 2, red line).
Again because of the higher average ionization, a somewhat larger mass is
required. The mass within $4000$\,\kms\ is now $\sim 1.3 \Msun$. The oxygen mass
is $\sim 1.0 \Msun$ and the carbon mass is $0.14 \Msun$. All other elements have
masses similar to those used for the model with the old code, except for Ca,
which is reduced to $\sim 0.013 \Msun$. The model now has $n_e \approx 4 \times
10^6$\,cm$^{-3}$ and $T_e \approx 3400$\,K. 

In the IR, the quality of the fits with the two versions of the code is very
similar (Figure 3, red line). As with the old code, a rather small mass of
silicon is sufficient to reproduce the [\SiI] line near $1.6 \mu$.

\section{Shell Models}

In this section we present shell models for nebular observations of SN\,2007gr
at 172 and 390 days after explosion. The models were computed using the updated
version of the code.  The results of one-zone models (in particular abundances
and their ratios) were used as the starting point for the shell models, which
were however optimised independently.  In contrast to the one-zone modelling we
do not present 'best fit' models for each epoch, but we try to find one model
which can reproduce the observations at 172 and 390 days simultaneously. Only
sodium and magnesium mass estimates change between the two epochs. 

While there is no IR spectrum at 172 days, there is an IR observation of SN
2007gr at 190 days after explosion. In order to be able to merge the optical and
the IR into a single spectrum, and lacking any IR photometry on day 172
\citep{Hunter09}, we extrapolated the 190-day IR observation to 172 days.  To do
this, we computed the IR-to-optical flux ratio of synthetic models at 172 and
190 days, and scaled the 190-day IR observations to reproduce the expected ratio
at 172 days. Since the difference is 18 days only, this extrapolation should be
reasonable. We can then compare the optical and IR flux at both 172 and 390
days. Shell models for the day 172 optical spectrum and the rescaled IR spectrum
of day 190 are shown in Figures 4 and 5, respectively, while fits for the day
390 optical and IR spectra are shown in Figures 6 and 7, respectively, as green
lines. 

The total mass of our model is $\sim 1.0 \Msun$ below 6000\,\kms\ (the model
becomes inaccurate at higher velocities because $\gamma$-ray deposition is very
small at low density). This is in rough agreement with the one-zone models (see
Tables 1 and 2). In general, the masses derived with the shell model are
slightly lower. This is expected since one-zone models usually produce too much
flux around the peak of the lines, which leads to an overestimate of the line
luminosity and thus of the amount of emitting material. Mass ratios of
individual elements are similar to the one-zone models. 

\begin{figure}
\includegraphics[width=89mm]{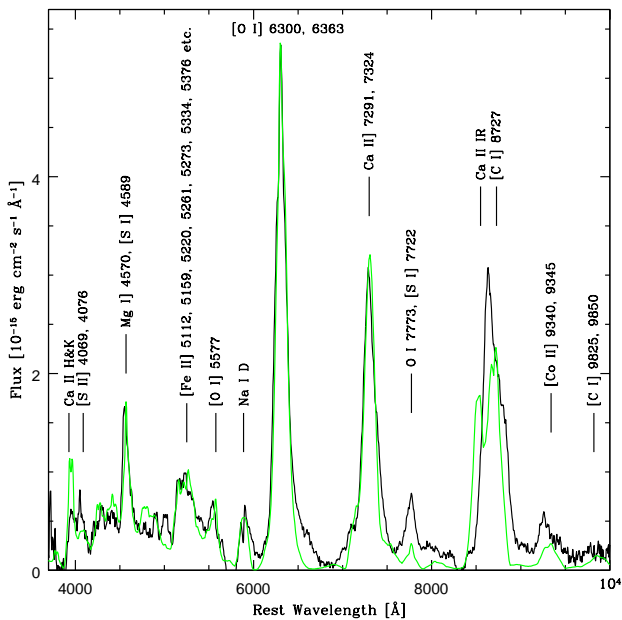}
\caption{SN\,2007gr: models for the optical spectrum at day 172 (29 Jan 2008,
black) with the shell version of the nebular code, which includes an 
improved treatment of ionization and oxygen recombination (green). }
\label{d172_shells_opt}
\end{figure}

\begin{figure}
\includegraphics[width=89mm]{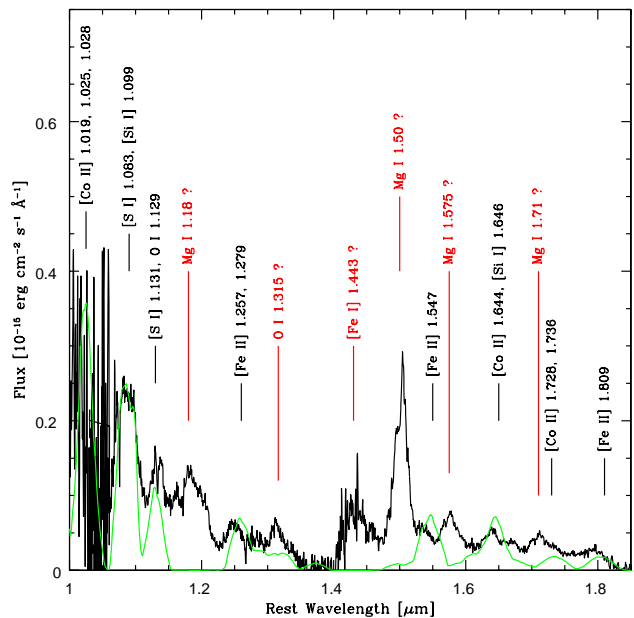}
\caption{SN\,2007gr: models for the IR spectrum at day 190, rescaled to day 
172 (16 Feb 2008, black) with the shell version of the nebular code, which
includes an improved treatment of ionization and oxygen recombination (green).
Lines marked in red are tentative identification of features that are not
reproduced in the current version of the code: they could be mostly 
recombination lines of \MgI\ and \OI. }
\label{d172_shells_opt}
\end{figure}

\begin{table*}
 \centering
 \begin{minipage}{140mm}
  \caption{Results of Shell Models. The masses listed here are located
  below 6000 km s$^{-1}$. About 0.4 M$_\odot$ of oxygen are expected
  above 6000 km s$^{-1}$. Since there is no information about the
  helium content of SN 2007gr the values for total mass are lower limits.}
  \begin{tabular}{cccccccccc}
  \hline
  epoch & Mass & M(C) & M(O) & M(Na) &  M(Mg) & M(Si) & M(S) & M(Ca) & M(\Nifs) \\
  days & $\Msun$& $\Msun$& $\Msun$& $\Msun$& $\Msun$& $\Msun$& $\Msun$& $\Msun$& $\Msun$ \\
 \hline
    172 & 1.02 & 0.15 & 0.76 & 5.0e-5 & 9.5e-4 & 0.005 & 0.020 & 0.009 & 0.076 \\
    390 & 1.05 & 0.15 & 0.76 & 5.0e-4 & 0.028  & 0.005 & 0.020 & 0.009 & 0.076 \\
\hline
\end{tabular}
\end{minipage}
\end{table*}

One drawback of the nebular modelling approach is the poorly constrained
distribution of \Nifs. The one-zone model assumes that \Nifs\ is  distributed
homogeneously. Some constraints on the precise distribution of \Nifs\ can be
derived from the profile of the \FeII\ blend near 5200\,\AA, but this is usually
weak in SNe\,Ib/c. In addition, not all \Nifs\ distributions can consistently
reproduce the evolution of the spectrum from 172 to 390 days. In our shell model
\Nifs\ is more centrally concentrated than in the one-zone models, in agreement
with theoretical expectations. The uncertainty on the total mass caused by this
assumption should be $\sim10$\%.

We use a clumping factor $\zeta = 2.5$, which gives the best overall agreement
with the observations, although individual lines or line ratios can sometimes be
better reproduced using different values. Any moderate value, $\zeta \lsim 10$
seems possible, taking into account uncertainties in the atomic data and the
calculation procedure.

Since the same model is used for both epochs (apart from differences in the Na
and Mg content), we discuss the two epochs together, focussing on individual
elements. 

The carbon distribution is mainly derived from modelling the [\CI] 8727\,\AA\
emission at 172 days. The flux in this region is a blend of [\CI] and \CaII\ 
emission and the observed flux profile cannot be reproduced by any model with
our nebular code. It may be that the \CaII\ flux is scattered by the carbon
line, a process which is not simulated by the code. Alternatively, [\FeII] lines
may be stronger than predicted by the model, as discussed in Section 4.2.1. At
day 172, we try to reproduce the flux at the wavelength of the [\CI] line. Using
the same carbon abundance, at 390 days the synthetic [\CI] 8727\,\AA\ line is
stronger than in the observations. This may be an indication that at day 172 the
line was powered by scattering of \CaII\ flux: since this process is not treated
in our code, we may have had to assume a higher carbon mass than is really
present in the SN ejecta. Since no other line of carbon is visible, the estimate
of the carbon mass may thus be uncertain by a factor of $\sim 2$. 

The oxygen distribution and mass are mainly derived from modelling the [\OI]
6300, 6363\,\AA\ doublet. The observations are matched rather well by the
synthetic flux at both epochs. The strength of the [\OI] 5577\,\AA\ line depends
on clumping. Using a clumping factor $\zeta = 2.5$ it is reproduced well at 172
days. At 390 days the line is no longer visible, since its formation is strongly
favoured at higher densities. With the new code, several permitted \OI\ lines
which arise from recombination can be modelled. The \OI\ 7773\,\AA\ line is not
reproduced with sufficient strength at 172 days, but at 390 days the synthetic
flux is in good agreement with the observations (see \citet{Maurer10b} for a
detailed discussion of the formation of this line). The \OI\ 8447\,\AA\ line is
blended with strong \CaII\ emission and cannot be identified. The \OI\,
9264\,\AA\ line is too weak at day 172. Since its formation is directly related
to the 7773\,\AA\ line this is expected. In the IR, the \OI\ 1.129\,$\mu$m line
is swamped by [\SI] 1.131\,$\mu$m. At 390 days the synthetic flux of the \OI\
7773\,\AA\ and 9264\,\AA\ lines reproduces the observations, taking into account
the possible background. The \OI\ 8447\,\AA\ line still cannot be identified in
the observations.

\begin{figure}
\includegraphics[width=89mm]{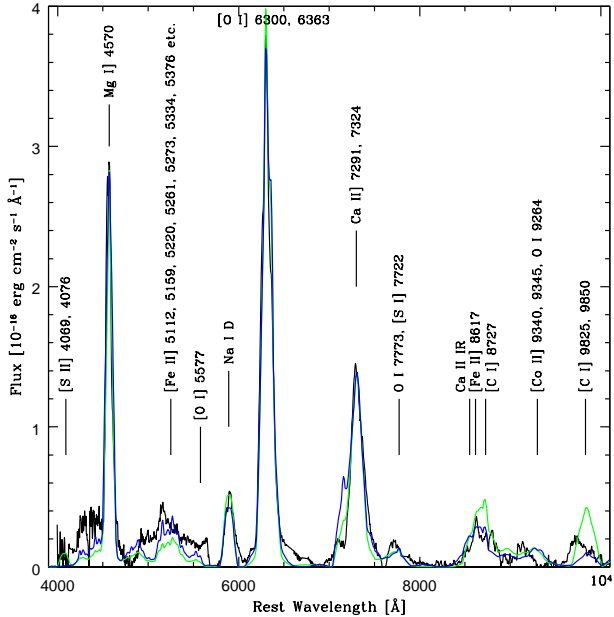}
\caption{SN\,2007gr: shell model for the optical spectrum at day 390 
(3 Sept 2008, black) with the shell version of the nebular code, which
includes an improved treatment of ionization and oxygen recombination. The  
green line shows a model computed for clumping $\zeta = 2.5$, the blue line is a
model computed for $\zeta = 66$. }
\label{d390_onezone_IR}
\end{figure}

\begin{figure}
\includegraphics[width=89mm]{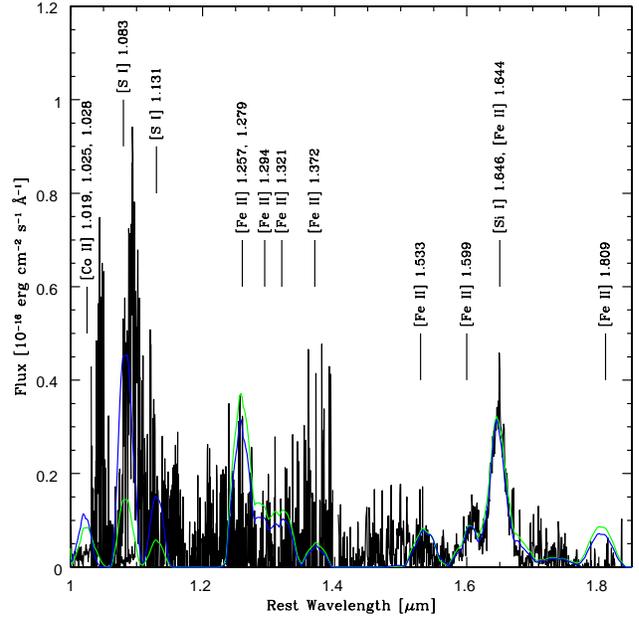}
\caption{SN\,2007gr: shell model for the IR spectrum at day 390 
(3 Sept 2008, black) with the shell version of the nebular code, which
includes an improved treatment of ionization and oxygen recombination . The  
green line shows a model computed for clumping $\zeta = 2.5$, the blue line is a
model computed for $\zeta = 66$. }
\label{d390_onezone_IR}
\end{figure}

Sodium and magnesium are modelled fitting the \NaI\,D and \MgI]\,4571\,\AA\
observations, but different masses are required at different epochs (see Sec.
4.2), which indicates that these ions are not accurately treated.  In fact, the
IR spectrum at day 190 shows a number of emission lines which are not reproduced
by the code (see red marks in Figure 5). The line at 1.5\,$\mu$ was identified
as \MgI\ 1.503\,$\mu$ by \citet{Pozzo06}. Several other lines can be identified
as permitted \MgI\ transitions, as indicated in Fig. 5. These lines could be
produced by recombination of \MgII. To our knowledge, this is the first time
these lines are identified in a nebular spectrum of a SN.  The flux in these
lines is intrinsically weak (note the factor $\sim 10$ difference in the
vertical scale of Fig. 5 with respect to Fig. 4). Inclusion of a detailed
treatment of \MgII\ recombination may result in a better fit of these lines and
in a more accurate determination of the Mg mass. 

Two other lines are not identified in the IR spectrum at day 190. The line at
$1.3\,\mu$m may be \OI\ 1.315\,$\mu$m. This would be part of the \OII\
recombination process. The emission near 1.43\,$\mu$m was identified as [\FeI]
1.443\,$\mu$m by \citet{Pozzo06} in the spectrum of a SN\,Ia obtained very early
in the nebular phase. While this identification may be possible, the lack of any
other [\FeI] lines, in particular 1.294 and 1.355\,$\mu$m, which arise from the
same lower level (the ground state of \FeI) and should have similar strength is
a strong argument against this option, which remains however the only one we
could find. 

The silicon distribution is modelled fitting the [\SiI] 1.588, 1.608 and
1.646\,$\mu$m lines at 390 days, when the [\SiI] feature can be observed
clearly. The line is blended with [\FeII] lines, and the same procedure
described in Section 4.1.2 is used to determine the mass of silicon. At 172 days
this feature is blended with [\CoII] emission. Also with the shell model the
silicon mass remains remarkably small.

The sulphur distribution can be obtained modelling the [\SII] 4070  line at 172
days, when it is more clearly visible, as well as the [\SI] 1.083 and
1.131\,$\mu$m lines. The [\SI] 7722\,\AA\ line is blended with \OI\ 7773\,\AA.
Since it is difficult to identify individual lines in the region around
4000\,\AA, we fix the sulphur mass by fitting the IR lines accurately, while
aiming at an acceptable agreement for the 4070\,\AA\ line. At 390 days the
synthetic IR lines seem to be too weak, but the observations are very noisy in
that spectral region. 

The calcium distribution and total mass are rather uncertain. The ratio of
\CaII] 7300\,\AA\ and the IR-triplet depends sensitively on density and
clumping. In addition, the IR-triplet may interact with \CI\ 8727\,\AA, a
process which is not simulated in the code. We fix the Ca mass modelling the
7300\,\AA\ emission at 172 days. This gives a consistent result also at day 390.

Since we cannot obtain an estimate of the helium mass, the total mass in
SN\,2007gr is uncertain. \citet{Hunter09} derive a total mass of $\sim$ 2
M$_\odot$ from light curve modelling, which would imply $\sim 0.5$\,M$\odot$ of
He.

Finally, the spectra for day 390 show an apparent mismatch between the optical
and IR Fe lines: most IR lines seem to be overestimated while the [\FeII]
emission near 5200\,\AA\ is too weak. In order to test whether this may be the
effect of clumping, we have computed a model with a clumping factor which brings
the optical synthetic spectrum into reasonable agreement with the observations.
This requires extreme clumping ($\zeta = 66$): the model is shown in Figures 6
and 7 as a blue line. Increased clumping leads also to a reduction in the flux
of the [\CI] lines in the optical. However, despite the large change in the flux
of the optical [\FeII] lines, the IR [\FeII] lines are almost unaffected:
increased clumping leads to a reduction of the population of the highly excited
levels from which the optical lines are formed, but leaves the much larger
population of the less excited levels from which the IR lines form basically
unchanged.  Another possibility would be to increase the \Nifs\ mass.  The
[\FeII] complex near 5200\,\AA\ can be reproduced in sufficient strength for
M(\Nifs)$ = 0.1 \Msun$, but then the IR Fe lines become too strong. Further
increase of the clumping seems unlikely: already with $\zeta = 66$ the spectrum
at day 172 is poorly reproduced. Considering the uncertainties in the background
subtraction, the Fe atomic data, the \Nifs\ distribution and the geometry, it is
not possible to determine the degree of clumping accurately, at least using the
code in its present state.

\section{Discussion}

The availability of a simultaneous optical-IR spectrum for SN\,2007gr makes it
possible to attempt an accurate estimate of the ejecta mass via a census of all
elements that are expected to be abundant in the ejecta of a SN\,Ic. 

Our models indicate that SN\,2007gr ejected $\sim 1 \Msun$ of heavy elements.
Our models are sensitive to material expanding with $v \lsim 6000$\,\kms.  We
adopt here the values from the shell model, since this offers the best
reproduction of the observed line fluxes and profiles.  We use a \Nifs\ mass of
$0.076 \Msun$, as determined from the luminosity at light curve peak
\citep{Valenti08,Hunter09}, and confirm that that value is a reasonable
estimate. We find that oxygen is the dominant constitutent of the heavy-element
ejecta, with a mass of $\sim 0.8 \Msun$.  The IME abundances, many of which can
be determined thanks to the IR spectrum, are quite small. In particular the Si
mass, which is determined via the [\SiI] 1.6\,$\mu$m line, is only $0.005
\Msun$. Since the Si mass is estimated from neutral silicon, which is the
dominant ion, this result is not much influenced by ionization. 

Our estimate of the Ca mass carries a larger uncertainty than for other elements
since it is estimated from lines of \CaII, which is less abundant than \CaI. 
Uncertainties in the ionization treatment can affect the mass estimate
significantly (a factor of 3 or so). 

Although we use moderate clumping ($\zeta = 2.5$), it is not possible, at the
current level of accuracy, to determine the exact degree of clumping. 

The rather low value of the ejected mass may be increased if the mass of He and
possibly Ne, which may be present to some extent, could also be accounted for.
\citet{Hunter09} estimated an ejected mass of $\sim$ 2 M$\odot$ from light curve
models. If this estimate is correct, the He mass of SN\,2007gr should be $\sim
0.5 - 1.0 \Msun$. This may be consistent with early-time observations of \HeI\
absorption, depending on the degree of mixing of He and radioactive material in
the outer layers of the SN ejecta.

The result that the mass of the CO core ejected by SN\,2007gr was $\sim 1 \Msun$
suggests that the SN progenitor was a star of comparatively low mass, possibly
similar to the progenitor of the prototypical low-energy SN\,Ic 1994I. In the
case of SN\,1994I a star of $\sim 15 \Msun$ was proposed as the likely
progenitor \citep{Nomoto94,Sauer06}. The star would have lost its outer hydrogen
and helium layers through binary interaction. The progenitor of SN\,2007gr may
have had a similar evolutionary history, with less stripping of the outer
layers. The estimate of the progenitor mass is also broadly consistent with the
observational constraints derived by \citet{Crockett08} from pre-explosion
images. 

\citet{Hunter09} estimated a \Nifs\ mass of 0.076 M$_\odot$. Using this value,
we can reproduce the temporal evolution of the nebular observations from 172 to
390 days. At 172 days the strength of the [\FeII] blend near 5200\,\AA\  is
reproduced reasonably well, while at 390 days the synthetic flux at these
wavelengths is too low. Uncertainties in a number of paramenters, like the
degree of clumping, the \Nifs\ distribution, Fe collision strengths, geometry,
reddening and background subtraction could be responsible for this.  A high
clumping factor ($\zeta \sim 60$) or a larger \Nifs\ mass ($\sim 0.1 \Msun$) can
approximately reproduce the strength of the [\FeII] blend. However, an increased
\Nifs\ mass leads to an over-estimate of less energetic [\FeII] lines redwards
of 6000\,\AA. Also, a \Nifs\ mass of more than $0.09 \Msun$ is inconsistent with
the 172 days spectrum. Therefore we estimate a total \Nifs\ mass between 0.07
and $0.09 \Msun$, but favour the lower values. Our results are in fact 
consistent with the estimates of the total and \Nifs\ mass presented in
\citet{Hunter09}. 

The elemental abundances that we derived are generally consistent with the
nucleosynthetic yields of a $\approx 15 \Msun$ star at roughly solar metallicity
\citep{WW95, ThiNomHash96, Nomoto97, Nomoto06}. The major discrepancy with
respect to those results is actually for silicon, which we find to be about one
order of magnitude less abundant. Our results are unlikely to be grossly in
error because of an incorrect estimate of the ionization: \SiI, the ion which is
observed in the IR and from which the Si mass is determined, is the dominant
silicon ion. On the other hand, possible uncertainties on the collision
strengths of \SiI, or of the \FeII\ lines which also contribute to the feature
at $1.65\,\mu$m from which the silicon abundance is measured, may cause
uncertainties. If the [\FeII] lines were completely removed, and the collision
strength for the [\SiI] lines remained the same, the silicon mass would increase
by a factor of $\sim 3$.  At the same time, the results cited above, while in
overall agreement, differ the most for IME (Si, S, Ca), and more recent
calculation (S.E. Woosley, priv. comm.) show a somewhat smaller silicon mass,
although again a factor of 10 error is unlikely. The most likely reason for the
discrepancy may lie with the collision strengths. A new calculation of these
values would be desireable. 

The detection of radio emission from SN\,2007gr, indicating the presence of
material moving at sub-relativistic velocities ($v \sim 0.6\,c$) in two opposite
jets \citep{Paragi10} is a very interesting finding. Normally, SNe associated
with relativistic outflows such as GRBs or XRFs have exceptionally large kinetic
energies and massive progenitors. This does not seem to be the case for
SN\,2007gr. Actually, the nebular spectra of SN\,2007gr show that the core of
the SN has rather low velocities, in agreement with the early phase, where
\citet{Valenti08} found maximum velocities of $\approx 10000$\,\kms. Actually,
in a paper that appeared after submission of this paper, \citet{Sod10} argue
that the radio data are consistent with a non-relativistic explosion, in line
with our optical results. 

Another peculiarity of SN\,2007gr was the detection of strong carbon lines at
early times \citep{Valenti08}. These lines were observed with velocities between
7000 and 10000\,\kms. In the nebular spectra the [\CI] 8727\,\AA\ line is
detected, suggesting the presence of carbon also at low velocities (below
5000\,\kms). Modelling indicates a C/O ratio of $\sim 0.1-0.2$, which is
consistent with the inner layers of the CO core of a massive star
\citep{Woosley94, ThiNomHash96, Nomoto97, Nomoto06, MaedMey03, MaedMey05}. 

Finally, GRB/SNe are thought to be very aspherical \citep{Maeda06}.  Some
evidence that all SNe\,Ib/c are aspherical, to some extent, was found by
\citet{Maeda08} using late-time spectra. \citet{Maurer10a} confirmed this, but
suggested that for SNe\,Ib/c which are not particularly energetic there may be
only a rather weak correlation between the velocities of the inner ejecta (``the
core") as derived from nebular spectroscopy and those of the outer ejecta, as
inferred from the spectra near maximum. The highest velocities detected in the
optical spectra ($v \sim 0.1\,c$) are always much lower than the
sub-relativistic velocities which can give rise to radio emission, so SN\,2007gr
may be a case where a very small amount of material was ejected at $v \sim
0.6\,c$ in an otherwise normal SN\,Ic \citep{Paragi10}. However, the recently
published results of \citet{Sod10}, who rule out material moving at $v > 0.2 c$
are in much better agreement with our results. From the nebular spectra we find
no evidence of major core asphericities in SN\,2007gr. This does not rule out
that some asphericity may affect the innermost regions of the ejecta
\citep{Maurer10a}. This may also help with the missing silicon if more fallback
occurs in the less nuclearly processed equatorial regions. Clearly, further
study of these very interesting events is necessary before a coherent picture
that encompasses the entire range of SNe\,Ib/c properties and their relation to
the those of the progenitor star can emerge. 

\section*{Acknowledgments} 

We gratefully acknowledge helpful conversations with
Stan Woosley and Ken Nomoto. Partial support from contract ASI/COFIS is
acknowledged.


\bsp

\label{lastpage}

\end{document}